\documentstyle[12pt]{article}
\begin{document}

\title{Delocalization of two interacting particles in a random potential:
one-dimensional electric interaction}
\author{J. C. Flores}
\date{Universidad de Tarapac\'a, Departamento de F\'\i sica, Casilla 7-D, Arica,
Chile}
\maketitle

\baselineskip=14pt

We consider a continuous one dimensional model of two charged interacting
particles in a random potential. The electric repulsion is strictly one
dimensional and it inhibits Anderson localization. In fact, the spectrum is
continuous. The case of electrical attraction is briefly studied and it
shows bounded states. So, the dynamics is sign-dependent for this model.We
support our analytical results with numerical simulations where the effect
of repulsion breaking localization is clearly observed.

$$
{} 
$$

A complete version of this paper (with figures): Phys.Rev.B {\bf 62}, 33
(2000).

\newpage\ 

The problem of one particle in a random potential has been studied widely
after the pioneer work by Anderson [1-5]. Now, it is accepted that
exponential localization could take place in such disordered systems
depending on the degree of disorder, the dimension $D$, and the energy of
the system. The role of internal spatial local correlations at the random
potential has been considered as mechanism of delocalization. In fact, there
is theoretical [6-19] and experimental [20] evidence of delocalization due
to correlations. Moreover, the role of no local correlation has been studied
showing new phenomena of delocalization [21]. Also, the dynamics of some
open random systems has been studied showing the breaking of Anderson
localization [22]. Recently, the role of interaction between particles in
disordered systems has been considered (two interacting particles or TIP).
Nevertheless, here some controversy exist [23-31]. It seems that for on-site
interaction (discrete model), the localization length is enhanced with
independence on the interaction sign.

In this paper, and for strictly one-dimensional electric repulsion, we show
that Anderson localization is broken for two interacting particles.
Moreover, the behavior of the system is not sign-independent. Our quantum
model is continuous and different of the usual discrete TIP model found in
the literature.

Consider Gauss's theorem in $D$ dimension, namely a charge $q$ producing an
electric field $E$ and a hypersphere of radius $r$ around the charge. The
theorem establishes that the electric flux through the hypersphere is
proportional to the charge enclosed. Namely,

\begin{equation}
Er^{D-1}\sim q. 
\end{equation}
So, the electric field produced by a charge is dependent on the dimension $D$%
. In three dimensions we have the usual $r^{-2}$ dependence and in strictly
one dimension the electric field is constant at the left and right of the
charge $q.$ Namely, in one dimension, the electric potential produced by one
charge $q$ is given by

\begin{equation}
V(x)=-\alpha q\left| x\right| ,\quad D=1, 
\end{equation}
where $\alpha $ is a proportionality constant and $x$ the position.

A remark concerning the one dimensional case. Since for a `sphere' in the
one dimension we have problems defining its surface, the result (2) is also
obtained by solving the Poisons equation $V^{\prime \prime }\sim q\delta (x)$
with the corresponding conditions of continuity $V^{+}(0)=V^{-}(0)$ , $\frac
d{dx}V^{+}(0)-\frac d{dx}V^{-}(0)\sim q$ and the symmetry condition $%
V(x)=V(-x)$.

In order to consider two charged interacting particles in strictly
one-dimensional disordered potential, we must consider the above electric
potential in the Hamiltonian: let $x$ and $y$ be the coordinate-operator of
two particles of charge $q$ and $Q$ , respectively; then the
Hamiltonian-operator of the systems is given by

\begin{equation}
H=-\frac 1{2m}\frac{\partial ^2}{\partial x^2}-\frac 1{2M}\frac{\partial ^2}{%
\partial y^2}+\xi (x)+\xi (y)-\alpha qQ\left| y-x\right| , 
\end{equation}
where $\xi $ denotes the random potential acting on every particle. To study
the above system we assume the following

$$
{} 
$$

(i) Particles do not overlap. Namely, we assume the charge $Q$ always on the
right of $q$.

(ii) Particles are distinguishable ($Q\neq q$), however by simplicity we
assume $M=m$.

(iii) The random potential is bounded ($\xi <\xi _{\max }$) and has no
internal correlation. Namely, if $\xi (x)=\sum f_i(x-x_i)$ then every
function $f_i$ is independent (also the random variable $x_i$). So, if there
are extended states they are not related to correlations at the random
potential [6-19].

(iv) No decoherence effects due to an external bath, or measurement, are
considered i.e. we have always a quantum pure state.

$$
$$

Let the state $\Psi _\omega (x,y)$ be solution of the time independent
Schr\"odinger equation, namely

\begin{equation}
H\Psi _\omega =\omega \Psi _\omega , 
\end{equation}
where $\omega $ is the energy ($\hbar =1$). The condition (i) requires that

\begin{equation}
\Psi _\omega (x,y)\left\{ 
\begin{array}{cc}
=0;y\leq x &  \\ 
\neq 0;y>x & 
\end{array}
\right. .
\end{equation}

Since the charge $Q$ is on the right of $q,$ the electrical interaction does
not require the absolute value i.e. $\left| y-x\right| =y-x.$ The
Schr\"odinger equation (4) becomes separable, and related to the solution of

\begin{equation}
-\frac 1{2m}\frac{\partial ^2}{\partial z^2}\phi +(\xi (z)\pm \alpha
qQz)\phi =\epsilon \phi , 
\end{equation}
where the constant $\epsilon $ is the energy. Since we are assuming
electrical repulsion ($Q>0,q>0$), the sign minus (or plus) in (6)
corresponds to the equation for $Q$ (or $q$). So, the charge $Q$, on the
right of $q$, experiences two forces: the random due to disorder, and the
electrical repulsion directed to the right. The charge $q$ is repelled to
the left.

The continuous Schr\"odinger equation (6) has been studied in the literature
and it corresponds to a particle in a disordered potential with a dc
electric field. For bounded disordered potential (iii), the states $\phi $
are extended in the direction of the field and the spectrum $\epsilon $ is
continuous [32-35]. This can be understood intuitively when we consider that
for a distance $\left| z\right| \gg \xi _{\max }/\alpha qQ$ the electric dc
field dominates the asymptotic behavior of the system and producing extended
states. We remark that our model is continuous, with bounded disorder, as
required by the theory of delocalization with electric dc field [32].

In this way, the solution of equation (4) can be written as

\begin{equation}
\Psi _\omega (x,y)=\left\{ \phi _\epsilon (x)\phi _{\omega -\epsilon
}(y)-\phi _\epsilon (y)\phi _{\omega -\epsilon }(x)\right\} ; x\leq
y, 
\end{equation}
where $\phi _\epsilon $ denotes the extended solution of energy $\epsilon $
of (6). We remark that the condition (i) (not overlap $\Psi (x,x)=0$) is
satisfied automatically from the antisymmetric condition for the wave
function. Then, for repulsion, localization is inhibited for two charged
interacting particles in strictly one dimension. Moreover, the spectrum $%
\omega $ is continuous and determined asymptotically by the electric
interaction. Namely, no periodic (or quasiperiodic) motion exists in the
system.

Numerical time-evolution calculations confirm our conjecture about
delocalization in the repulsive case: Figure (1a) shows the time-dispersion
for two wavepackets (particles) when interaction and disorder do not exist.
The initial condition is the localized state $\Psi (x,y,t=0)=\delta
_{x,x_o}\delta _{y.y_o}$ and the regime is ballistic. It shows the
probability $\int dx\left| \Psi (x,y,t)\right| ^2$ and $\int dy\left| \Psi
(x,y,t)\right| ^2$ to find the particle $Q$ and $q$ respectively at time $%
t\neq 0$. Figure (1b) shows the two particles under one-dimensional electric
repulsion and the same initial conditions of (1a). Remark the time
separation of the center of mass for every particle because repulsion.
Figure (1c) shows the evolution when only disorder is present, and one notes
the absence of dispersion for the wavepackets. Finally, Figure (1d) shows
the combined effect of repulsion and localization. Clearly, disorder does
not stop the separation due to repulsion of both particles. The numerical
calculations were carried-out by solving the Schr\"odinger equation related
to (3) with a simple procedure (finite-differences). The potential $\xi $
was constructed using a random number generator. The spatial boundary was
avoided by considering a finite number of iterations. This number was found
by using the iteration procedure related to figure (1b), the most fast
motion. All figures have the same parameters (number of iterations, step,
disorder, etc.).

On the other hand, when electrical attraction is considered in our
one-dimensional systems, the numerical results show the existence of bounded
states. Figures (2a) and (2b) show the temporal evolution for two
interacting particles in a random potential. (2a) shows the repulsive
behavior related to delocalization. Figure (2b) shows the evolution when one
changes the parameter of repulsion by attraction and keeping the same
initial conditions and disorder ($qQ\rightarrow -qQ$). Clearly the behavior
between repulsion and attraction is different. Since we are assuming
attraction between both particles, the charge $Q$ is pushed by the field to
the left against $q$; but from (i) the wave function vanishes when $y=x$.
From an analytical point of view, our result can be understood when we
consider no disorder and the usual coordinate change to the centre of mass.
Namely $2X=x+y$ and the relative coordinate $2r=x-y$. The non overlap
condition at $r=0$ and the constant attractive electric field gives origin
to bounded states [36]. The finite disorder ($\xi <\xi _{\max }$) does not
change the behavior of the system governed asymptotically by the electric
field. Discrete spectrum and bounded states exist in the system in this case.

Finally, we notice that the one-dimensional electric case studied here is
formally similar to two infinite parallel planes separated by a distance $%
\left| y-x\right| $ where the electric force is constant. This tells us that
plane charged molecules are good candidates to test our result. Also
quantum-wire, with one-dimensional directed electric-flux, seems an
interesting candidate to explore our conjectures.

In conclusion: we have presented theoretical evidence that strict one
dimensional electrical repulsion, which is long-range, breaks Anderson
localization for two interacting charged particles. This result is different
to the discrete on-site interaction which only enhances the localization
length, but does not break localization. In our case, the attraction and
repulsion behavior are sign-depending. These results were confirmed
numerically.

I thank Professor S. N. Evangelou who explain us the TIP\ Problem
(International Workshop on Disordered Systems with Correlated-Disorder,
Arica'98, PELICAN and FDI-UTA\ Projects). Interesting comments were
furnished by Professor Chumin Wang Chen, Professor V. Bellani (FDI-UTA and
PELICAN Projects) and Professor E. Lazo (UTA) . This work was partially
completed at Pavia University (CICOPS Scholarship), and partially supported
by FONDECYT (proyecto 1000439).

\newpage\ 

FIGURE CAPTION

$$
{} 
$$

Figure 1. (1a) Time evolution probability for two particles without disorder
and electric interaction. The initial condition is $\Psi (x,y)=\delta
_{x,x_0}\delta _{y.y_o}$. The dispersion is ballistic. For simplicity we do
not consider antisymmetric states. (1b) Time evolution probability with
electric field repulsion. The parameters are the same as in figure (1a).
Still we have no disorder. Remark the increasing separation between the
wavepackets with time. (1c) The same situation as in figure (1a) but with
disorder and without electric interaction. Clearly, diffusion does not exist
because Anderson localization. (1d) Combined influence of electric repulsion
and disorder. Disorder does not stop the repulsion process and
delocalization operates.

$$
{} 
$$

Figure 2. (2a) shows the repulsion for two electric charges in a random
potential. Figure (2b) shows the attractive case with the same parameter of
disorder and initial conditions. Clearly the behavior is different in both
cases.

\end{document}